\documentclass[twocolumn,showpacs,aps,floatfix,superscriptaddress]{revtex4}
\usepackage{amsmath}
\usepackage{graphicx}
\begin{document}
\title{Dynamics of Multi-Player Games}
\author{E.~Ben-Naim}\email{ebn@lanl.gov} 
\affiliation{Theoretical Division and Center for Nonlinear Studies,
Los Alamos National Laboratory, Los Alamos, New Mexico 87545 USA}
\author{B.~Kahng}\email{kahng@phya.snu.ac.kr} 
\affiliation{Theoretical Division and Center for Nonlinear Studies,
Los Alamos National Laboratory, Los Alamos, New Mexico 87545 USA}
\affiliation{School of Physics and Astronomy and Center for
Theoretical Physics, Seoul National University, Seoul 151-747, Korea}
\author{J.~S.~Kim}\email{nakzii@phya.snu.ac.kr} 
\affiliation{Theoretical Division and Center for Nonlinear Studies,
Los Alamos National Laboratory, Los Alamos, New Mexico 87545 USA}
\affiliation{School of Physics and Astronomy and Center for
Theoretical Physics, Seoul National University, Seoul 151-747, Korea}
\begin{abstract}
We analyze the dynamics of competitions with a large number of
players. In our model, $n$ players compete against each other and the
winner is decided based on the standings: in each competition, the
$m$th ranked player wins. We solve for the long time limit of the
distribution of the number of wins for all $n$ and $m$ and find three
different scenarios. When the best player wins, the standings are most
competitive as there is one-tier with a clear differentiation between
strong and weak players. When an intermediate player wins, the
standings are two-tier with equally-strong players in the top tier and
clearly-separated players in the lower tier. When the worst player
wins, the standings are least competitive as there is one tier in
which all of the players are equal. This behavior is understood via
scaling analysis of the nonlinear evolution equations.
\end{abstract}
\maketitle
\section{Introduction}
Interacting particle or agent-based techniques are a central method in
the physics of complex systems. This methodology heavily relies on the
dynamics of the agents or the interactions between the agents, as
defined on a microscopic level \cite{ww}. In this respect, this
approach is orthogonal to the traditional game theoretic framework
that is based on the global utility or function of the system, as
defined on a macroscopic level \cite{ft}.

Such physics-inspired approaches, where agents are treated as
particles in a physical system, have recently led to quantitative
predictions in a wide variety of social and economic systems
\cite{hfv,ckfl,bvr}. Current areas of interest include the
distribution of income and wealth \cite{ikr,dy,fs,sr}, opinion
dynamics \cite{wdan,smo,bkr}, the propagation of innovation and ideas
\cite{dz}, and the emergence of social hierarchies
\cite{btd,ss,msk,br}.

In the latter example, most relevant to this study, competition is the
mechanism responsible for the emergence of disparate social classes in
human and animal communities. A recently introduced competition
process \cite{btd,br} is based on two-player competitions where the
stronger player wins with a fixed probability and the weaker player
wins with a smaller probability \cite{bvr1}.  This theory has proved
to be useful for understanding major team sports and for analysis of
game results data \cite{bvr}.

In this study, we consider multi-player games and address the
situation where the outcome of a game is completely deterministic.
In our model, a large number of players $n$ participate in the
game, and in each competition, the $m$th ranked player always
wins. The number of wins measures the strength of a player.
Furthermore, the distribution of the number of wins characterizes
the nature of the standings. We address the time-evolution of this
distribution using the rate equation approach, and then, solve for
the long-time asymptotic behavior using scaling techniques.

Our main result is that there are three types of standings. When the
best player wins, $m=1$, there is a clear notion of player strength;
the higher the ranking the larger the winning rate. When an
intermediate player wins, $1<m<n$, the standings have two tiers.
Players in the lower tier are well separated, but players in the
upper-tier are all equally strong. When the weakest player wins,
$m=n$, the lower tier disappears and all of the players are equal in
strength. In this sense, when the best player wins, the environment is
most competitive, and when the worst player wins it is the least
competitive.

The rest of this paper is organized as follows. We introduce the
model in section II. In Section III, we analyze in detail
three-player competitions, addressing situations where the best,
intermediate, and worst player wins, in order. We then consider
games with an arbitrary number of players and pay special
attention to the large-$n$ limit in Section IV. We conclude in
section V.

\section{The multi-player model}

\begin{figure}[t]
\vspace*{0.cm} \includegraphics*[width=0.37\textwidth]{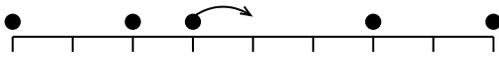}
\caption{Illustration of the multiplayer game with
 $n=5$ and $m=3$.} \label{ill}
\end{figure}

Our system consists of $N$ players that compete against each other. In
each competition $n$ players are randomly chosen from the total pool
of players. The winner is decided based upon the ranking: the $m$th
ranked player always wins the game [Fig.~\ref{ill}]. Let $k_i$ be the
number of wins of the $i$th ranked player in the competition, i.e.,
\hbox{$k_1\geq \cdots \geq k_m\geq \cdots\geq k_n$}, then
\begin{equation}
\label{rule} (k_1,\ldots,k_m,\ldots k_n)\to
(k_1,\ldots,k_m+1,\ldots,k_n).
\end{equation}
Tie-breakers are decided by a coin-toss, i.e., when two or more
players are tied, their relative ranking is determined in a
completely random fashion. Initially, players start with no wins,
$k=0$.

These competition rules are relevant in a wide variety of contexts. In
sports competitions, the strongest player often emerges as the
winner. In social contexts and especially in politics, being a
centrist often pays off, and furthermore, there are auctions where the
second highest bidder wins. Finally, identifying wins with financial
assets, the situation where the weakest player wins mimics a strong
welfare system where the rich support the poor.

We set the competition rate such that the number of competitions in a
unit time equals the total number of players. Thence, each player
participates in $n$ games per unit time, and furthermore, the average
number of wins $\langle k\rangle$ simply equals time
\begin{equation}
\label{kav} \langle k\rangle =t.
\end{equation}
At large times, it is natural to analyze the winning rate, that is,
the number of wins normalized by time, $x=k/t$. Similarly, from
our definition of the competition rate, the average winning rate
equals one
\begin{equation}
\label{xav} \langle x\rangle =1.
\end{equation}

Our goal is to characterize how the number of wins, or alternatively,
the winning rate are distributed in the long time limit. We note that
since the players are randomly chosen in each competition, the number
of games played by a given player is a fluctuating
quantity. Nevertheless, since this process is completely random,
fluctuations in the number of games played by a given player scale as
the square-root of time, and thus, these fluctuations become
irrelevant in the long time limit. Also, we consider the thermodynamic
limit, $N\to\infty$.

\section{Three player games}

We first analyze the three player case, $n=3$, because it nicely
demonstrates the full spectrum of possibilities. We detail the
three scenarios where the best, intermediate, and worst, players
win in order.

\subsection{Best player wins}

Let us first analyze the case where the best player wins. That is, if
the number of wins of the three players are $k_1\geq k_2\geq k_3$,
then the game outcome is as follows
\begin{equation}
\label{best}
(k_1,k_2,k_3)\to (k_1+1,k_2,k_3).
\end{equation}

Let $f_k(t)$ be the probability distribution of players with $k\geq 0$
wins at time $t$. This distribution is properly normalized, $\sum_k
f_k=1$, and it evolves according to the nonlinear
difference-differential equation
\begin{eqnarray}
\label{re-f-a}
\frac{df_k}{dt}&=&{3\choose 1}(f_{k-1}F^2_{k-1}-f_k\,F^2_k)\\
&+&{3\choose 2}\left(f^2_{k-1}F_{k-1}-f^2_k\,F_k\right)\nonumber\\
&+&{3\choose 3}\left(f^3_{k-1}-f^3_k\right).\nonumber
\end{eqnarray}
Here, we used the cumulative distributions $F_k=\sum_{j=0}^{k-1}f_j$
and $G_k=\sum_{j=k+1}^\infty f_j$ of players with fitness smaller than
and larger than $k$, respectively. The two cumulative distributions
are of course related, $F_k+G_{k-1}=1$. The first pair of terms
accounts for games where it is unambiguous who the top player is. The
next pair accounts for two-way ties for first, and the last pair for three
way ties. Each pair of terms contains a gain term and a loss term that
differ by a simple index shift. The binomial coefficients account for
the number of distinct ways there are to choose the players. For
example, there are ${3\choose 1}=3$ ways to choose the top player in
the first case. This master equation should be solved subject to the
initial condition $f_k(0)=\delta_{k,0}$ and the boundary condition
$f_{-1}(t)=0$.  One can verify by summing the equations that the total
probability is conserved $\frac{d}{dt}\sum_k f_k=0$, and that the
average fitness $\langle k\rangle =\sum_k k f_k$ evolves as in
(\ref{kav}), $d\langle k\rangle/dt=1$.

For theoretical analysis, it is convenient to study the cumulative
distribution $F_k$.  Summing the rate equations (\ref{re-f-a}), we
obtain closed equations for the cumulative distribution
\begin{eqnarray}
\label{re-F-a}
\frac{dF_k}{dt}&=&-3(F_k-F_{k-1})F^2_{k-1}\\
&&-3(F_k-F_{k-1})^2F_{k-1}\nonumber\\
&&-(F_k-F_{k-1})^3.\nonumber
\end{eqnarray}
Here, we used $f_k=F_{k+1}-F_k$. This master equation is subject
to the initial condition $F_k(0)=1$ and the boundary condition
$F_{-1}(t)=0$.

We are interested in the long time limit. Since the number of wins is
expected to grow linearly with time, $k\sim t$, we may treat the
number of wins as a continuous variable, $F_{k-1}=F_k-\frac{\partial
F}{\partial k}+\frac{1}{2}\frac{\partial^2 F}{\partial k^2}+\cdots$.
Asymptotically, since $\frac{\partial F}{\partial k}\propto t^{-1}$
and $\frac{\partial^2F}{\partial k^2}\propto t^{-2}$, etc., second-
and higher-order terms become negligible compared with the first order
terms. To leading order, the cumulative distribution obeys the
following partial differential equation
\begin{equation}
\label{F-eq-a}
\frac{\partial F}{\partial t}+3F^2\frac{\partial
F}{\partial k}=0.
\end{equation}
From dimensional analysis of this equation, we anticipate that the
cumulative distribution obeys the scaling form
\begin{equation}
\label{Phi-def} F_k(t)\simeq \Phi(k/t)
\end{equation}
with the boundary conditions $\Phi(0)=0$ and $\Phi(\infty)=1$. In
other words, instead of concentrating on the number of wins $k$, we
focus on the winning rate $x=k/t$. In the long time limit, the
cumulative distribution of winning rates $\Phi(x)$ becomes
stationary. Of course, the actual distribution of winning rates
$\phi(x)$ also becomes stationary, and it is related to the
distribution of the number of wins by the scaling transformation
\begin{equation}
\label{phi-def} f_k(t)\simeq t^{-1}\phi(k/t)
\end{equation}
with $\phi(x)=\Phi'(x)$. Since the average winning rate equals one
(\ref{xav}), the distribution of winning rates must satisfy
\begin{equation}
\label{constraint}
1=\int _0^{\infty} dx \, x \, \Phi'(x).
\end{equation}

\begin{figure}[t]
\vspace*{0.cm} \includegraphics*[width=0.35\textwidth]{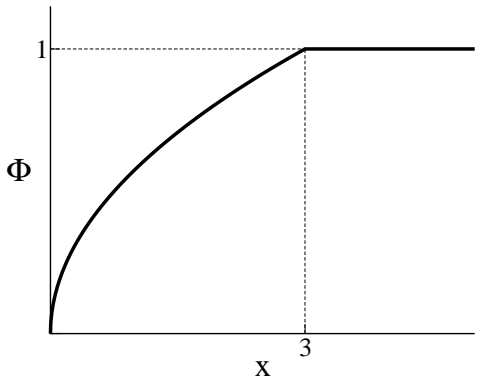}
\caption{The stationary distribution of winning rates
(\ref{Phi-sol-a}) for the case $n=3$, $m=1$.} \label{n3m1-fig}
\end{figure}

Substituting the definition (\ref{Phi-def}) into the master
equation (\ref{F-eq-a}), the stationary distribution satisfies
\begin{equation}
\label{Phi-eq-a} \Phi'(x)[3\Phi^2-x]=0.
\end{equation}
There are two solutions: (i) The constant solution, $\Phi(x)={\rm
const}$, and (ii) The algebraic solution $\Phi_s(x)=(x/3)^{1/3}$.
Invoking the boundary condition $\lim_{x\to\infty}\Phi(x)=1$ we find
[Fig.~\ref{n3m1-fig}]
\begin{equation}
\label{Phi-sol-a} \Phi(x)=
\begin{cases}
(x/3)^{1/2}&x\leq 3\\
1&x\geq 3.\\
\end{cases}
\end{equation}
One can verify that this stationary distribution satisfies the
constraint (\ref{constraint}) so that the average winning rate
equals one. This result generalizes the linear stationary
distribution found for two player games \cite{br}.

\begin{figure}[t]
\vspace*{0.cm} \includegraphics*[width=0.4\textwidth]{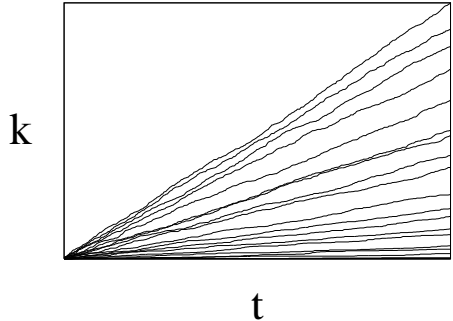}
\caption{Schematic illustration of the number of wins $k$ versus
time $t$ when the best player wins. Shown are results of
simulations with 20 players.} \label{kta}
\end{figure}

Initially, all the players are identical, but by the random
competition process, some players end up at the top of the standings
and some at the bottom. This directly follows from the fact that the
distribution of winning rates is nontrivial.  Also, since $\phi(x)\sim
x^{-1/2}$ as $x\to 0$, the distribution of winning-rate is nonuniform
and there are many more players with very low winning rates.  When the
number of players is finite, a clear ranking emerges, and every player
wins at a different rate.  Moreover, after a transient regime, the
rankings do not change with time [Fig.~\ref{kta}].

We note that in our scaling analysis, situations where there is a two-
or three-way tie for first do not contribute. This is the case because
the number of wins grows linearly with time and therefore, the
probability of finding two players with the same number of wins can be
neglected. Such terms do affect how the distribution of the number of
wins approaches a stationary form, but they do not affect the final
form of the stationary distribution.

\subsection{Intermediate player wins}

Next, we address the case where the intermediate player wins,
\begin{equation}
\label{intermediate}
(k_1,k_2,k_3)\to (k_1,k_2+1,k_3).
\end{equation}
Now, there are four terms in the master equation
\begin{eqnarray}
\label{re-f-b}
\frac{df_k}{dt}&=&{3\choose 1}{2\choose 1}(f_{k-1}F_{k-1}G_{k-1}-f_kF_kG_k)\\
&+&{3\choose 1}\left(f^2_{k-1}G_{k-1}-f^2_k\,G_k\right)\nonumber\\
&+&{3\choose 2}\left(f^2_{k-1}F_{k-1}-f^2_k\,F_k\right)\nonumber\\
&+&{3\choose 3}\left(f^3_{k-1}-f^3_k\right).\nonumber
\end{eqnarray}
The first pair of terms accounts for situations where there are no
ties and then the combinatorial prefactor is a product of the number
of ways to choose the intermediate player times the number of ways
to choose the best player. The next two pairs of terms account for
situations where there is a two-way tie for best and worst,
respectively. Again, the last pair of terms accounts for three-way
ties. These equations conserve the total probability, $\sum_k f_k
=1$, and they are also consistent with (\ref{kav}).

\begin{figure}[t]
\vspace*{0.cm} \includegraphics*[width=0.35\textwidth]{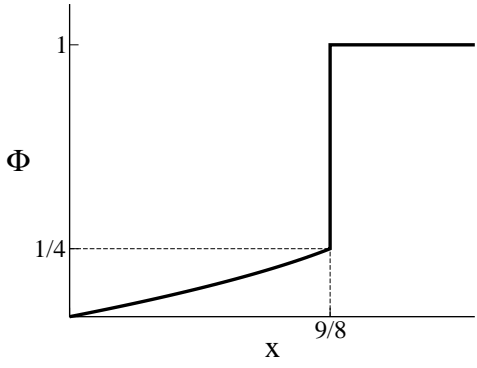}
\caption{The stationary distribution of winning rates
(\ref{Phi-sol-b}) for $n=3$, $m=2$.} \label{n3m2-fig}
\end{figure}

Summing the rate equations
(\ref{re-f-b}), we obtain closed equations for the cumulative
distribution
\begin{eqnarray}
\label{re-F-b}
\frac{dF_k}{dt}&=&-6(F_k-F_{k-1})F_{k-1}G_{k-1}\\
&&-3(F_k-F_{k-1})^2(F_{k-1}+G_{k-1})\nonumber\\
&&-(F_k-F_{k-1})^3.\nonumber
\end{eqnarray}
For clarity, we use both of the cumulative distributions, but note
that this equation is definitely closed in $F_k$ because of the
relation $G_k=1-F_{k+1}$. Taking the continuum limit and keeping only
first-order derivatives, the cumulative distribution obeys the
following partial differential equation \hbox{$\frac{\partial
F}{\partial t}+6F(1-F)\frac{\partial F}{\partial k}=0$} with the
boundary conditions $F_0=0$ and $\lim_{k\to\infty}F_k=1$.
Substituting the definition of the stationary distribution of winning
rates (\ref{Phi-def}) into this partial differential equation, we
arrive at
\begin{equation}
\label{Phi-eq-b}
\Phi'(x)[6\Phi(1-\Phi)-x]=0,
\end{equation}
an equation that is subject to the boundary conditions $\Phi(0)=0$ and
$\lim_{x\to\infty}\Phi(x)=1$.  There are two solutions: (i) The
constant solution, $\Phi(x)={\rm const}$, and (ii) The root of the
second-order polynomial
$\Phi_s(x)=\frac{1}{2}\big(1-\sqrt{1-2x/3}\big)$.  Invoking the
boundary conditions, we conclude [Fig.~\ref{n3m2-fig}]
\begin{equation}
\label{Phi-sol-b}
\Phi(x)=
\begin{cases}
\frac{1}{2}\left(1-\sqrt{1-\frac{2}{3}x}\right)&x<x_0\\
1&x>x_0.\\
\end{cases}
\end{equation}
As the nontrivial solution is bounded $\Phi_s(x)\leq 1/2$, the
cumulative distribution must have a discontinuity. We have
implicitly assumed that this discontinuity is located at
$x_0<3/2$.

The location of this discontinuity is dictated by the average number
of wins constraint. Substituting the stationary distribution
(\ref{Phi-sol-b}) into (\ref{constraint}) then
\begin{eqnarray*}
1=\int_0^{x_0} dx\, x \,\Phi'(x)+x_0[1-\Phi(x_0)].
\end{eqnarray*}
In writing this equality, we utilized the fact that the stationary
distribution has a discontinuity at $x_0$ and that the size of this
discontinuity is $1-\Phi_0$.  Integrating by parts, we obtain an
implicit equation for the location of the discontinuity
\begin{equation}
\label{x0-eq}
1=x_0-\int_0^{x_0}\, dx\, \Phi(x).
\end{equation}
Substituting the stationary solution (\ref{Phi-sol-b}) into this
equation and performing the integration, we find after several
manipulations that the location of the singularity satisfies the
cubic equation $x_0^2\big(x_0-\frac{9}{8}\big)=0$. The location of
the discontinuity is therefore
\begin{equation}
\label{x0-sol}
x_0=\frac{9}{8}.
\end{equation}
This completes the solution (\ref{Phi-sol-b}) for the scaling function. The
size of the discontinuity follows from $\Phi_0\equiv \Phi(x_0)=1/4$.

\begin{figure}[t]
\vspace*{0.cm} \includegraphics*[width=0.4\textwidth]{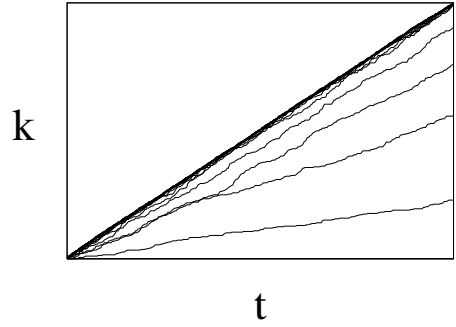}
\caption{Schematic illustration of the number of wins $k$ versus
time $t$ when the intermediate player wins. Shown are results of
simulations with 20 players.} \label{ktb}
\end{figure}

There is an alternative way to find the location of the
discontinuity. Let us transform the integration over $x$ into an
integration over $\Phi$ using the equality
\begin{eqnarray}
x_0\Phi_0&=&\int_0^{x_0}\, dx\, \Phi(x)+\int_0^{\Phi_0} d\Phi\, x(\Phi).
\end{eqnarray}
This transforms the equation for the location of the discontinuity
(\ref{x0-eq}) into an equation for the size of the jump
\begin{equation}
\label{Phi0-eq}
1=x_0(1-\Phi_0)+\int_0^{\Phi_0}\, d\Phi\, x(\Phi).
\end{equation}
Substituting $x(\Phi)=6\Phi(1-\Phi)$ we arrive at the cubic equation
for the variable $\Phi_0$, $1=6\Phi_0-9\Phi_0^2+4\Phi_0^3$. The
relevant solution is $\Phi_0=\frac{1}{4}$, from which we conclude
$x_0=9/8$. For three-player games, there is no particular advantage
for either of the two approaches: both (\ref{x0-eq}) and (\ref{Phi0-eq})
involve cubic polynomials.  However, in general, the latter approach
is superior because it does not require an explicit solution for
$\Phi(x)$.

The scaling function corresponding to the win-number distribution is
therefore
\begin{eqnarray*}
\phi(x)=\frac{1}{6}\,\left({1-\frac{2}{3}\,x\,}\right)^{-1/2}\,+
\,\frac{3}{4}\,\delta\left(x-\frac{9}{8}\right),
\end{eqnarray*}
where $\delta(x)$ denotes the Kronecker delta function. The
win-number distribution contains two components. The first is a
nontrivial distribution of players with winning rate $x<x_0$ and
the second reflects that a finite fraction of the players have the
maximal winning rate $x=x_0$. Thus, the standings have a two-tier
structure. Players in the lower tier have different strengths and
there is a clear differentiation among them [Fig.~\ref{ktb}].
Players in the upper-tier are essentially equal in strength as
they all win with the same rate.  A fraction $\Phi_0=\frac{1}{4}$
belongs to the lower tier and a complementary fraction
$1-\Phi_0=\frac{3}{4}$ belongs to the upper tier. Interestingly,
the upper-tier has the form of a condensate. We note that a
condensate, located at the bottom, rather than at the top as is
the case here, was found in the diversity model in Ref.~\cite{br}.

\subsection{Worst player wins}

\begin{figure}[t]
\vspace*{0.cm} \includegraphics*[width=0.35\textwidth]{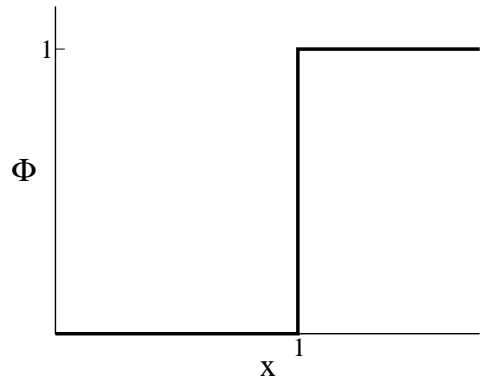}
\caption{The stationary distribution of winning rates
(\ref{Phi-sol-c}) for the case $n=m=3$.} \label{n3m3-fig}
\end{figure}

Last, we address the case where the worst player wins \cite{bvr1,ktnr} 
\begin{equation}
(k_1,k_2,k_3)\to (k_1,k_2,k_3+1).
\end{equation}
Here, the distribution of the number of wins evolves according to
\begin{eqnarray}
\label{re-f-c}
\frac{df_k}{dt}&=&{3\choose 1}(f_{k-1}G^2_{k-1}-f_k\,G^2_k)\\
&+&{3\choose 2}\left(f^2_{k-1}G_{k-1}-f^2_k\,G_k\right)\nonumber\\
&+&{3\choose 3}\left(f^3_{k-1}-f^3_k\right).\nonumber
\end{eqnarray}
This equation is obtained from (\ref{re-f-a}) simply by replacing
the cumulative distribution $F_k$ with $G_k$. The closed equation
for the cumulative distribution is now
\begin{eqnarray}
\label{re-F-c}
\frac{dF_k}{dt}&=&-3(F_k-F_{k-1})G^2_{k-1}\\
&&-3(F_k-F_{k-1})^2G_{k-1}\nonumber\\
&&-(F_k-F_{k-1})^3.\nonumber
\end{eqnarray}
In the continuum limit, this equation becomes \hbox{$\frac{\partial
F}{\partial t}+3(1-F)^2\frac{\partial F}{\partial k}=0$}, and consequently,
the stationary distribution satisfies
\begin{equation}
\label{Phi-eq-c}
\Phi'(x)[3(1-\Phi)^2-x]=0.
\end{equation}
Now, there is only one solution, the constant $\Phi(x)={\rm
const}$, and because of the boundary conditions $\Phi(0)=0$ and
$\lim_{x\to\infty}\Phi(x)=1$, the stationary distribution is a
step function: $\Phi(x)=1$ for $x > x_0$ and $\Phi(x)=0$ for
$x<x_0$. In other words, $\Phi(x)=\Theta(x-x_0)$. Substituting
this form into the condition (\ref{constraint}), the location of
the discontinuity is simply $x_0=1$, and therefore
[Fig.~\ref{n3m3-fig}]
\begin{equation}
\label{Phi-sol-c}
\Phi(x)=\Theta(x-1)
\end{equation}
where $\Theta(x)$ is the Heaviside step function.  When the worst
player wins, the standings no longer contain a lower-tier: they
consist only of an upper-tier where all players have the same
winning rate, $\phi(x)=\delta(x-1)$.

\section{Arbitrary number of players}

Let us now consider the most general case where there are $n$
players and the $m$th ranked player wins as in (\ref{rule}). It is
straightforward to generalize the rate equations for the
cumulative distribution. Repeating the scaling analysis above,
Eqs.~(\ref{Phi-eq-a}) and (\ref{Phi-eq-b}) for the stationary
distribution (\ref{Phi-def}) generalize as follows:
\begin{equation}
\label{Phi-eq-d} \Phi'(x)[C\Phi^{n-m}(1-\Phi)^{m-1}-x]=0.
\end{equation}
The constant $C$ equals the number of ways to choose the $m$th
ranked player times the number of ways to choose the $m-1$ higher
ranked players
\begin{equation}
\label{constant} C={n\choose 1}{n-1\choose
m-1}=\frac{n!}{(n-m)!(m-1)!}.
\end{equation}
Again, there are two solutions: (i) The constant solution,
$\Phi'(x)=0$, and (ii) The root of the $(n-1)$th-order polynomial
\begin{equation}
\label{Phi-eq-e} C\Phi^{n-m}(1-\Phi)^{m-1}=x.
\end{equation}
We now analyze the three cases where the best, an intermediate, and
the worst player win, in order.

\noindent{\bf Best player wins ($m=1$):} In this case, the
stationary distribution can be calculated analytically,
\begin{equation}
\label{Phi-sol-d}
\Phi(x)=
\begin{cases}
(x/n)^{1/(n-1)}&x\leq n;\\
1&x\geq n.
\end{cases}
\end{equation}
One can verify that this solution is consistent with (\ref{xav}). We
see that in general, when the best player wins there is no
discontinuity and $\Phi_0=1$. As for three-player games, the standings
consist of a single tier where some players rank high and some rank
low. Also, the winning rate of the top players equals the number of
players, $x_0=n$.  In general, the distribution of the number of wins
is algebraic.

\begin{figure}[t]
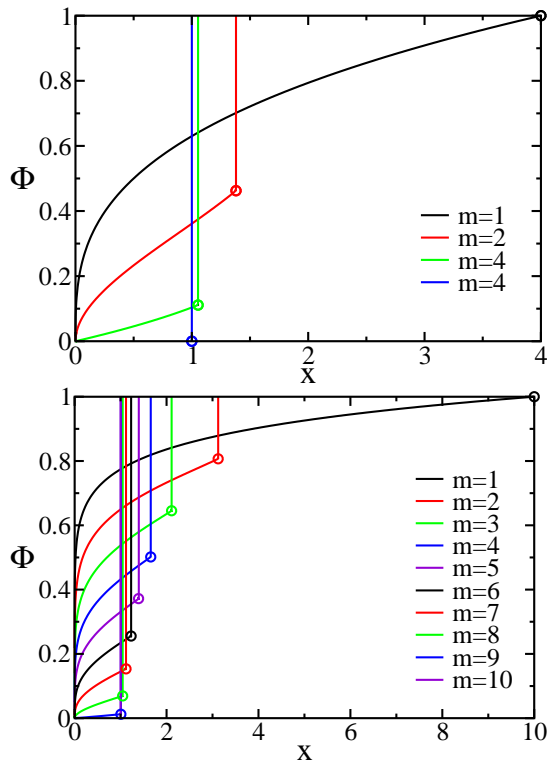

\vspace*{0.cm} \includegraphics*[width=0.4\textwidth]{fig7a.eps}
\vspace*{0.cm} \includegraphics*[width=0.4\textwidth]{fig7b.eps}
\caption{The stationary distribution of winning rates $\Phi(x)$ for
$n=4$ (top) and $n=10$ (bottom). Shown are Monte Carlo simulation
results with $N=10^6$ particles at time $t=10^5$. The circles are the
theoretical predictions for the maximal winning rate $x_0$ and the
size of the lower tier $\Phi_0$.}
\label{n4n10-fig}
\end{figure}

\noindent{\bf Intermediate player wins ($1< m <n$):} Based on the
behavior for three player games, we expect
\begin{equation}
\label{Phi-sol-e} \Phi(x)=
\begin{cases}
\Phi_s(x)&x<x_0;\\
1&x\geq x_0.
\end{cases}
\end{equation}
Here, $\Phi_s(x)$ is the solution of (\ref{Phi-eq-e}). Numerical
simulations confirm this behavior [Fig.~\ref{n4n10-fig}]. Thus, we
conclude that in general, there are two tiers. In the upper tier,
all players have the same winning rate, while in the lower tier
different players win at different rates. Generally, a finite
fraction $\Phi_0$ belongs to the lower tier and the complementary
fraction $1-\Phi_0$ belongs to the upper tier.

Our Monte Carlo simulations are performed by simply mimicking the
competition process. The system consists of a large number of players
$N$, all starting with no wins. In each elemental step, $n$ players
are chosen and ranked and the $m$th ranked player is awarded a win
(tied players are ranked in a random fashion). Time is augmented by
$1/N$ after each such step. This elemental step is then repeated.

The parameters $x_0$ and $\Phi_0$ characterize two important
properties: the maximal winning rate and the size of each tier.  Thus,
we focus on the behavior of these two parameters and pay special
attention to the large-$n$ limit. Substituting the stationary
distribution (\ref{Phi-sol-e}) into the constraint (\ref{constraint}),
the maximal winning rate $x_0$ follows from the very same
Eq.~(\ref{x0-eq}). Similarly, the size of the lower tier follows from
Eq.~(\ref{Phi0-eq}). In this case, the latter is a polynomial of
degree $n+1$, so numerically, one solves first for $\Phi_0$ and then
uses (\ref{Phi-eq-e}) to obtain $x_0$. We verified these theoretical
predictions for the cases $n=4$ and $n=10$ using Monte Carlo
simulations [Fig.~\ref{n4n10-fig}].

For completeness, we mention that it is possible to rewrite
Eq.~(\ref{Phi0-eq}) in a compact form. Using the definition of the
Beta function
\begin{eqnarray}
\label{relation}
\int_0^1 d\Phi\, \Phi^{n-m}(1-\Phi)^{m-1}&=&B(n-m+1,m)\\
&=&\frac{(n-m)!(m-1)!}{n!}\nonumber\\
&=&C^{-1}\nonumber
\end{eqnarray}
we relate the definite integral above with the combinatorial
constant in (\ref{constant}). Substituting the governing equation
for the stationary distribution (\ref{Phi-eq-e}) into the equation
for the size of the lower-tier (\ref{Phi0-eq}) gives
\begin{equation}
C^{-1}=\Phi_0^{n-m}(1-\Phi_0)^m+\int_0^{\Phi_0} d\Phi\,
\Phi^{n-m}(1-\Phi)^{m-1}.
\end{equation}
Using the relation (\ref{relation}), we arrive at a convenient
equation for the size of the lower tier $\Phi_0$
\begin{equation}
\int_{\Phi_0}^1
d\Phi\,\Phi^{n-m}(1-\Phi)^{m-1}=\Phi_0^{n-m}(1-\Phi_0)^m.
\end{equation}
This is a polynomial of degree $n+1$.

\begin{figure}[t]
\vspace*{0.cm} \includegraphics*[width=0.35\textwidth]{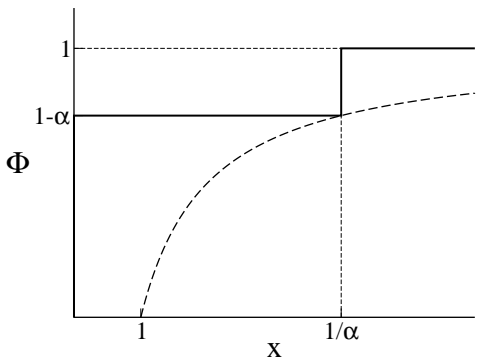}
\caption{The infinite-$n$ limit. From Eq.~(\ref{limit}), the points
$(x_0,\Phi_0)$ all lie on the curve $\Phi=(x-1)/x$.} \label{limit-fig}
\end{figure}

Let us consider the limit $n\to\infty$ and $m\to\infty$ with the ratio
$\alpha=m/n$ kept constant. For example, the case $\alpha=1/2$
corresponds to the situation where the median player is the winner. To
solve the governing equation for the stationary distribution in the
large-$n$ limit, we estimate the combinatorial constant $C$ using
Eq.~(\ref{constant}) and the Stirling formula $n!\sim (2\pi
n)^{1/2}(n/e)^n$.  Eq.~(\ref{Phi-eq-e}) becomes
\begin{equation}
\sqrt{\frac{n\alpha}{2\pi(1-\alpha)}}\left(\frac{\Phi}{1-\alpha}\right)^{n-m}
\left(\frac{1-\Phi}{\alpha}\right)^{m-1}\sim x.
\end{equation}
Taking the power $1/n$ on both sides of this equation, and then
the limit $n\to\infty$, we arrive at the very simple equation,
\begin{equation}
\Big(\frac{\Phi}{1-\alpha}\Big)^{1-\alpha}\Big(\frac{1-\Phi}{\alpha}\Big)^{\alpha}
=1.
\end{equation}
By inspection, the solution is constant, $\Phi=1-\alpha$. Using
$\Phi_0=1-\alpha$ and employing the condition $\langle x\rangle
=1$ yields the location of the condensate
\begin{equation}
\label{limit}
x_0=1/\alpha, \qquad \Phi_0=1-\alpha.
\end{equation}
This result is consistent with the expected behaviors $x_0\to
\infty$ as $\alpha\to 0$ and $x_0(\alpha=1)=1$ (see the worst
player wins discussion below). Therefore, the stationary
distribution contains two steps when the number of players
participating in each game diverges [Fig.~\ref{limit-fig}]
\begin{equation}
\label{Phi-infinite}
\Phi(x)=
\begin{cases}
0& x<0 \\
1-\alpha&0<x<\alpha^{-1} \\
1&\alpha^{-1}<x.
\end{cases}
\end{equation}
The stationary distribution corresponding to the number of wins
therefore consists of two delta-functions:
\hbox{$\phi(x)=(1-\alpha)\delta(x)+\alpha\delta(x-1/\alpha)$}.  Thus,
as the number of players participating in a game grows, the winning
rate of players in the lower tier diminishes, and eventually, they
become indistinguishable.

For example, for $n=10$, the quantity $\Phi_0$ is roughly linear in
$\alpha$ and the maximal winning rate $x_0$ is roughly proportional to
$\alpha^{-1}$ [Fig.~\ref{n4n10-fig}]. Nevertheless, for moderate $n$
there are still significant deviations from the limiting asymptotic
behavior. A refined asymptotic analysis shows that
$\Phi_0-(1-\alpha)\sim \sqrt{\alpha(1-\alpha)\ln n/n}$ and that
$x_0\simeq (1-\Phi_0)^{-1}$ \cite{pk}.  Therefore, the convergence is
slow and nonuniform (i.e., $\alpha$-dependent). Despite the slow
convergence, the infinite-$n$ limit is very instructive as it shows
that the structure of the lower-tier becomes trivial as the number of
players in a game becomes very large. It also shows that the size of
the jump becomes proportional to the rank of the winning player.

It is also possible to analytically obtain the stationary
distribution in the limit of small winning rates, $x\to 0$. Since
the cumulative distribution is small, $\Phi\to 0$, the governing
equation (\ref{Phi-eq-e}) can be approximated by $C\Phi^{n-m}=x$.
As a result, the cumulative distribution vanishes algebraically
\begin{equation}
\Phi(x)\sim x^{\frac{1}{n-m}},
\end{equation}
as $x\to 0$. This behavior holds as long as $m<n$.

\noindent{\bf Worst player wins ($m=n$):} In this case, the roots of
the polynomial (\ref{Phi-eq-e}) are not physical because they
correspond to either monotonically increasing solutions or they are
larger than unity. Thus, the only solution is a constant and following
the same reasoning as above we conclude that the stationary
distribution is the step function (\ref{Phi-sol-c}). Again, the upper
tier disappears and all players have the same winning rate. In other
words, there is very strong parity.

We note that while the winning rate of all players approaches the same
value, there are still small differences between players.  Based on
the behavior for two-player games, we expect that the distribution of
the number of wins follows a traveling wave form $F_k(t)\to U(k-t)$ as
$t\to\infty$ \cite{bvr}. As the differences among the players are
small, the ranking continually evolves with time. Such analysis is
beyond the scope of the approach above. Nevertheless, the dependence
on the number of players may be quite interesting.

Let us imagine that wins represent wealth. Then, the strong players are
the rich and the the weak players are the poor.  Competitions in which
the weakest player wins mimic a strong welfare mechanism where the
poor benefits from interactions with the rich. In such a scenario,
social inequalities are small.

\section{Conclusions}

In conclusion, we have studied multi-player games where the winner is
decided deterministically based upon the ranking. We focused on the
long time limit where situations with two or more tied players are
generally irrelevant. We analyzed the stationary distribution of
winning rates using scaling analysis of the nonlinear master
equations.

The shape of the stationary distribution reflects three qualitatively
different types of behavior. When the best player wins, there are
clear differences between the players as they advance at different
rates. When an intermediate player wins, the standings are organized
into two tiers. The upper tier has the form of a condensate with all
of the top players winning at the same rate; in contrast, the lower
tier players win at different rates. Interestingly, the same
qualitative behavior emerges when the second player wins as when the
second to last player wins. When the worst player wins, all of the
players are equal in strength.

The behavior in the limit of an infinite number of players greatly
simplifies. In this limit, the change from upper tier only standings
to lower tier only standings occurs in a continuous fashion.
Moreover, the size of the upper tier is simply proportional to the
rank of the winner while the maximal winning rate is inversely
proportional to this parameter.

In the context of sports competitions, these results are consistent
with our intuition. We view standings that clearly differentiate the
players as a competitive environment. Then, having the best player win
results in the most competitive environment, while having the worst
player win leads to the least competitive environment.  As the rank of
the winning player is varied from best to worst, the environment is
gradually changed from highly competitive to non-competitive.  This is
the case because the size of the competitive tier decreases as the
strength of the winning player declines.

In the context of social dynamics, these results have very clear
implications: they suggest that a welfare strategy that aims to
eliminate social hierarchies must be based on supporting the very
poor as all players become equal when the weakest benefits from
competitions.

Our asymptotic analysis focuses on the most basic characteristic,
the winning rate. However, there are interesting questions that
may be asked when tiers of equal-strength players emerge. For
example, the structure of the upper tier can be further explored
by characterizing relative fluctuations in the strengths of the
top players. Similarly, the dynamical evolution of the ranking
when all players are equally strong may be interesting as well.

\acknowledgments{We thank Paul Krapivsky for analysis of the large-$n$
limit. We acknowledge financial support from DOE grant W-7405-ENG-36
and KRF Grant R14-2002-059-010000-0 of the ABRL program funded by the
Korean government (MOEHRD).}

\end{document}